# A Spatial Analysis of Disposable Income in Ireland: A GWR Approach

Paul Kilgarriff *, Martin Charlton **

* Luxembourg Institute for Socio Economic Research (LISER)

** National Centre for Geocomputation, Maynooth University

**Abstract**

This paper examines the spatial distribution of income in Ireland. Median gross household disposable income data from the CSO, available at the Electoral Division (ED) level, is used to explore the spatial variability in income. Geary's C highlights the spatial dependence of income, highlighting that the distribution of income is not random across space and is influenced by location. Given the presence of spatial autocorrelation, utilising a global OLS regression will lead to biased results. Geographically Weighted Regression (GWR) is used to examine the spatial heterogeneity of income and the impact of local demographic drivers on income. GWR results show the demographic drivers have varying levels of influence on income across locations. Lone parent has a stronger negative impact in the Cork commuter belt than it does in the Dublin commuter belt. The relationship between household income and the demographic context of the area is a complicated one. This paper attempts to examine these relationships acknowledging the impact of space.

**Introduction**

Previous attempts to estimate disposable income at the local level in Ireland used spatial microsimulation methods (Ballas et al., 2005; O'Donoghue, Ballas, et al., 2013). These methods relied upon CSO county level income data (CSO, 2017a), EU-SILC survey data (CSO, 2017b) and Census data (CSO, 2016) to simulate disposable income at the household and electoral division (ED) level (Kilgarriff et al., 2016; O'Donoghue, Farrell, et al., 2013). Alternatively measures of deprivation such as the Pobal Deprivation Index (PDI) (Haase, 2017) were used to examine levels of affluence or deprivation at a small area level. The index is calculated based on demographic, social class and labour indicators from the SAPS which give each small area index scores with a mean of zero and standard deviation of ten (Haase and Pratschke, 2012). These measures were useful given the lack of published income data at a local level, however there were limitations with both as they are attempted proxies of a true measure of income or poverty.

In Summer 2019, the central statistics office (CSO) released for the first time data on household disposable income, medical card holders and dependence on social welfare payments at a small spatial scale, the Electoral Division (ED) level of which there are 3,409 (CSO, 2019a). The personal income register (PIR) (derived from Revenue Commissioners and Department of Employment Affairs and Social Protection data) is linked with the Census of Population Analysis dataset (COPA) achieving a 95% match (CSO, 2020). The resulting internal CSO dataset makes it possible to calculate a measure of median income per ED.

Previous analysis of simulated disposable income highlighted the regional disparity in income between urban and rural and Dublin vs other cities (Kilgarriff et al., 2016). The PDI highlighted the differences in deprivation levels between urban and rural areas and within urban areas (Haase, 2017). For many socio-economic and demographic measures differences exist across space. Explanations for regional differences can be explained by the existence of agglomeration economies (Glaeser, 2011; Glaeser and Gottlieb, 2009; Rosenthal and Strange, 2003). Urban areas benefit from dense labour markets (Combes et al., 2008), cultural diversity (Ottaviano and Peri, 2005) a high-skilled workforce (Glaeser and Resseger, 2010) and knowledge spillovers (Rosenthal and Strange, 2006) among others.

Although measures such as the Pobal Deprivation Index (Haase, 2017) can be used as a proxy classifying an area affluent or deprived, these new data offer to opportunity to test whether these indices are closely correlated and secondly examine how income specifically relates to other variables and indicators such as unemployment, housing tenure and education. This income information combined with census variables can help in illustrating the differences in income between areas and how these differences relate to differences between other socio-economic variables.

*Spatial Dependence*

Differences between areas lead us onto the issue of spatial dependence or spatial autocorrelation. Spatial dependence is a lack of independence between observations and is emphasized by the importance of relative location. There is a relationship between what happens at one location and what happens at another (Anselin, 2013). This is expressed by Toblers (Tobler, 1970) first law of geography, *"everything is related to everything else, but near things are more related than distant things"*. Spatial dependence can result from measurement errors due in part to the way in which spatial units are aggregated which may not reflect the spatial scope of the data. This can lead to spill over of errors across boundaries. Choosing variables based on aggregates at a sub-regional level, assumes that the choice of these regions reflects the spatial patterns (Bivand et al., 2008). Another cause is related to a variety of spatial interactions, these interactions are a function of distance and space (Anselin, 2013). The creation of areal units relate to the issue of the Modifiable Areal Unit Problem (MAUP), which states that statistical measures are sensitive to the way in which data is aggregated (Openshaw, 1984).

From the CSO income data we are already aware of the areas with highest/lowest levels of disposable income, we add to this information by exploring the spatial variability of income. A random forest algorithm (Ho, 1995; Liaw and Wiener, 2002) is used for model selection. A Global OLS model is performed. Proceeding the global OLS model we utilise Geary's C (Geary, 1954) to test for the presence of spatial autocorrelation. Given the nature of income, we expect income to be exhibit spatial autocorrelation, pattern of income not to be random. Geographically weighted regression (Fotheringham et al., 1998) is used to examine spatial heterogeneity in the model. Additionally multi-scale geographically weighted regression (MS-GWR) and geographically weighted ridge regression (GWRR) (Wheeler, 2007) are explored to control for local multicollinearity effects. This enables us to test the local relationship between income and some of its drivers.

**Methodology**

The first task involves selecting appropriate independent variables to use in our model. The objective of the analysis is to examine what the main drivers of median gross household disposable income (MGHDI) are. A random forest model (Liaw and Wiener, 2002) is used to measure the power each variable has in predicting income. The bias of the OLS regression coefficients as a result of spatial autocorrelation is highlighted and identified using Geary's C. A spatial error model is then utilised to estimate our model accounting for the spatial autocorrelation. To explore the heterogeneity in data, Geographically Weighted Regression (GWR) is used. GWR enables us to explore the difference in the relationship between the dependent and independent variables across space.

*Model specification*

Model specification is one of the most important steps involved in undertaking spatial regression. An incorrectly specified model may lead to biases in the estimators. The first step involves selecting the independent variables. A test is then performed to ensure our model does not have issues such as multicollinearity. However as our regression is spatial, multicollinearity may not be present in the global model but present at the local scale. Additional tests must be performed before we can settle on a final selection of independent variables to be used in the model.

*Variable Selection*

After assembling the SAPS data for 2016 (CSO, 2016) along with the CSO Income in Ireland 2016 data (CSO, 2019b) additional variables are derived from SAPS variables. The variables used were derived from previous UK studies which classified Census districts (Charlton et al., 1985; Webber, 1977, 1979). A list of similar variables available in the Irish Census were created by Brunsdon, Charlton and Rigby (2018) to cluster Census small areas. The reproducible code produced by Brunsdon, Rigby and Charlton (2014) is used as part of this analysis. These variables include age categories, household composition, housing tenure, employment status, education level along with other measures related to services and health.

Overall, the dataset contains over 100 variables. In order to select a more manageable set of independent variables some researchers have chosen variables adopting a "sledgehammer approach" (Mather and Openshaw, 1974). Variables maybe selected using social and economic theory, such as age and employment being strong predictors of income, however a data driven method is used to inform our choice of variables. The random forest algorithm (Liaw and Wiener, 2002) is utilised to examine variable "importance" in relation to the dependent variable - median gross household disposable income MGHDI. Our original dataset is split into two samples, training and validation (also called out-of-bag (OOB) (Breiman, 1996)). The decision tree is trained on the training data sample before being predicted on the validation or OOB data sample. The variable importance measured as percentage increase in mean square error (MSE) enables us to rank and compare variables against each other. Variables with a % increase MSE >10 are chosen to proceed to the next step. Unfortunately the random forest algorithm does not tells us whether the variables are correlated so further tests are required. This step allows the data to drive the decision process around variable selection. Variables are not chosen arbitrarily.

After a cross correlation matrix of the variables, highly correlated variables were dropped, for example EU national and born outside of Ireland, lone parent and separated and pensioner and age group 65+. Variables, which are highly correlated, with more than one variable are first to be dropped. To decide between two variables, the variable with the higher % increase in MSE from the random forest was chosen. The variance inflation factor (VIF) is a method used to test for multicollinearity, as a general rule a VIF above five suggests multicollinearity (Harris and

Jarvis, 2014). Rooms per household was dropped due to a high VIF. Whereas people per room was dropped due to a high coefficient and standard error in our OLS model. Table 1 shows the VIF scores for the variables with MGHDI the dependent variable.

**Table 1: Variance Inflation Factor of variables**

| Age 65+ | Born outside Ireland | Single Person | Lone Parent | Two car hh | Unemployed | Commerce | Internet | No education | High education |
|---|---|---|---|---|---|---|---|---|---|
| 1.72 | 2.10 | 2.52 | 3.04 | 5.72 | 2.57 | 1.73 | 3.04 | 2.78 | 2.84 |

Due to the high VIF factor for two car household (5.72), the variable is dropped from the analysis. After the variable selection process, we are left with nine independent variables detailed in table 2 which do not exhibit multicollinearity in the global model. Histograms of the OLS variables are also shown in figure 1.

**Table 2: Final Variables Global Model**

| Variable | Detail |
|---|---|
| Median Gross Household Disposable Income | Household |
| Age 65+ | % of persons |
| Born outside Ireland | % of persons |
| Single Person | % of persons |
| Lone Parent | % of families |
| Unemployed | % of persons over 15 |
| Commerce workers | % of persons at work |
| Internet (Broadband) | % of households |
| No formal education | % of person over 15 |
| Education level higher bachelor's degree and higher | % of persons over 15 |

**Figure 1: Histograms of OLS variables**

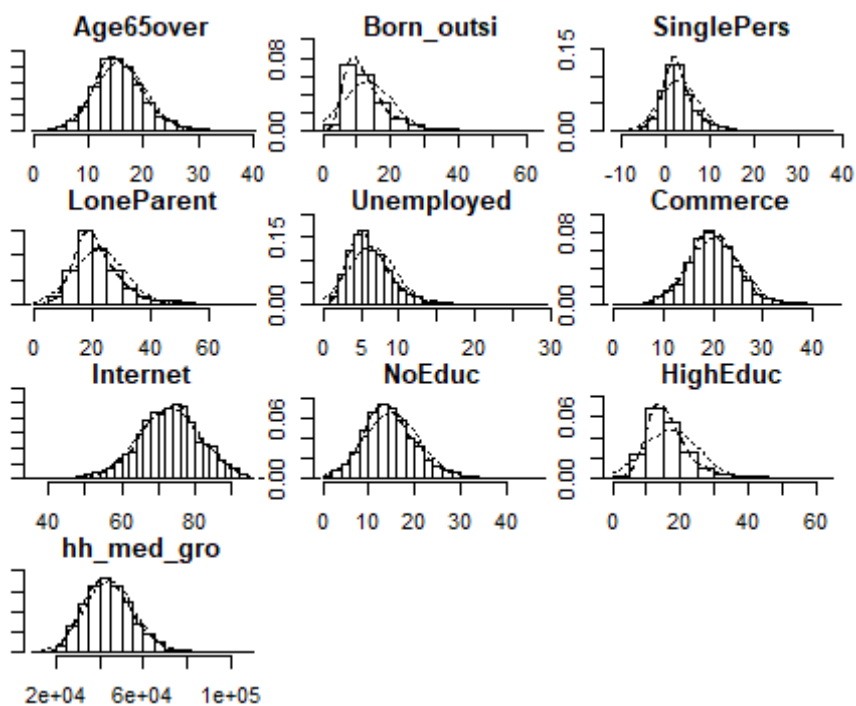

If we assume that our data is spatially independent, OLS regression may be used to examine the relationship between our variables. If however there is a degree of spatial dependence in our model, this will bias our estimates. In other words, there is a component in the error term which is not random and which can be explained by distance and space. The results from the OLS regression can be termed global, as no allowance for spatial heterogeneity is made.

**Table 3: OLS Regression Results**

```
===============================================================
                                   Dependent variable:
                        ---------------------------------------
                        Median Household Gross Disposable Income
---------------------------------------------------------------
Age 65+                              -530.835***
                                       (22.897)
Born outside Ireland                 -465.286***
                                       (15.601)
Single Person                        -410.649***
                                       (29.124)
Lone Parent                          -149.255***
                                       (13.341)
Unemployed                           -482.034***
                                       (44.663)
Commerce workers                      282.039***
                                       (20.848)
Internet (Broadband)                  194.030***
                                       (17.701)
Low Education                         -67.732***
                                       (23.688)
High Education                        582.044***
                                       (17.636)
Constant                           37,736.720***
                                     (1,452.110)
---------------------------------------------------------------
Observations                            3,409
R2                                      0.800
Adjusted R2                             0.800
Residual Std. Error                5,103.329 (df = 3399)
F Statistic                     1,512.802*** (df = 9; 3399)
===============================================================
Note:                           *p<0.1; **p<0.05; ***p<0.01
```

From table 3, we can see the sign of the coefficients are as expected. A higher share of households employed in commerce, high levels of education and high levels of internet are associated with higher levels of income. These results can be viewed as a first step and give some indication as to what the main drivers of income are.

*Spatial Dependence*

An initial test of spatial dependence is to group EDs by CSO urban-rural classification (CSO, 2019c). If spatial dependence exists in the data, we cannot treat the EDs as independent observations for a conventional regression. OLS regressions provide global estimate of the parameters in the model – that is, the relationship is considered to be uniform across the study area, as given in equation 1:

$$y_i = \beta_0 + \sum_k \beta_k \, x_{ik} + \varepsilon_i \quad (Equation\ 1)$$

If spatial dependence exists in our data, any spatial pattern in the data is transferred to the OLS residuals ($\varepsilon_i$) – these are no longer independent and violate the OLS assumption of BLUE.

From figure 2 we can see that income is heterogeneous across Ireland. Even within counties, there is a lot of variation. The areas with the highest median value is located in Dublin city and Cork rural with a high level of urban influence. These areas with a high urban influence are the commuter zones. It would appear that households in Cork with high levels of income choose to live in these commuter zones as opposed to living in the city. In Dublin this is not the case as high incomes households seem to live in the city. There could be a variety of reasons for this such as levels of amenities in the city, quality of housing and commuting times.

**Figure 2: Boxplots of Median Income per ED by Urban-Rural classification**

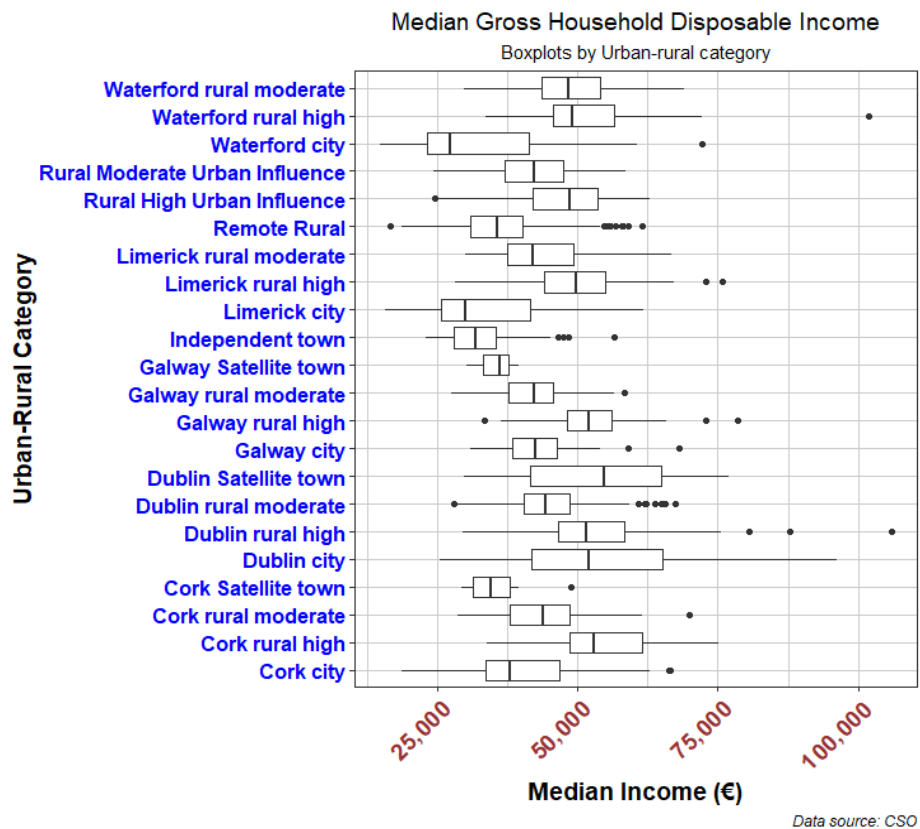

If our data is spatially autocorrelated then our error term will be correlated with the dependent variable, in other words there is a spatial dependence in the data which is biasing the results.

*Geary's C – Spatial Autocorrelation*

Geary's C is a measure of spatial autocorrelation developed by Roy C. Geary (Geary, 1954). Unlike autocorrelation, spatial autocorrelation considers how observations influence each other through a network such as the map of EDs (De Jong et al., 1984). The formula for Geary's C can be expressed as the ratio between the sum of square differences between locations i and j (two ED locations) on the numerator and the sum of square deviations from the mean on the denominator (Anselin, 2019). As Geary's C examines differences which are more relevant to the local, as opposed to products in Moran's I (the most widely used measure of spatial autocorrelation) (Unwin, 1996). Technology constraints made calculating Geary's C difficult (Jeffers, 1973) which may have reduced its usage. The measure of Geary's C utilised here comes from the spdep package in r (Bivand and Wong, 2018). Table 4 shows the results of the Geary's C test, with no spatial autocorrelation a value of 1 is expected. Any value below 1 indicates positive spatial autocorrelation. For calculating Geary's C k-nn of 23 was used. This was optimal bandwidth of the GWR-Basic using the AIC approach, which included all variables. What is clear from table 4 is the reduction in spatial autocorrelation of the residuals using GWR. Without an adjustment spatial autocorrelation is present in both the dependent (Median Household Gross Disposable Income) and independent variables. This is a further justification for using GWR over an OLS or Spatial Error Model. Geary's C is a global measure of spatial autocorrelation and requires further interpretation and investigation (Unwin, 1996). Although, Anselin (1995) has produced a local measures including Geary's C, by decomposing the global statistics. Local Geary is utilised to identify spatial clusters, areas with high values located close to areas with high values. Local Geary can be performed using the open software GeoDa (Anselin et al., 2010).

**Table 4: Geary's C values of variables**

| Variable | Geary's C |
| --- | --- |
| Median Household Gross Disposable Income | 0.51*** |
| Single Person | 0.76*** |
| High Education | 0.43*** |
| Age 65+ | 0.69*** |
| No Education | 0.61*** |
| Born outside of Ireland | 0.63*** |
| Commerce | 0.47*** |
| Residuals OLS | 0.56*** |
| Residuals Spatial Error Model | 0.56*** |
| Residuals GWR Basic (all variables) | 0.98*** |
| Residuals GWR Basic (Model 1) | 0.94*** |
| Residuals LCR-GWR (Model 2) | 1.00*** |

*Spatial error model*

A spatial lag model or spatial error model can be used to understand the nature of the spatial dependence found using Geary's C. These model require a spatial weights matrix. This weights matrix is constructed using binary contiguity between spatial units. If two spatial units in a grid are neighbours a value of 1 is assigned, otherwise a zero (Anselin, 2013). The measures of spatial autocorrelation Moran's I (Moran, 1948) and Geary's C (Geary, 1954) used binary contiguity. Spatial weights matrix using queen contiguity receives its name from the queen in chess, an area is considered a neighbour if they touch in any direction. It is at this stage that we remove islands as they will have zero neighbours. In the case of an irregular grid such as census areal units, the need to row standardise weights emerges, so that over importance is not given to units with many neighbours (Bivand et al., 2008). A spatial lag model attempts to relate the value of a variable at one location, with the value of that variable at another location:

$$y_{t-k} = L^k y$$

For a global OLS regression we make the assumption that our dependent variable covary between fixed locations adhering to a random distribution. Due to the way in which observations are aggregated, there will be spillovers across boundaries making locations covary with each other, near things are more related to each other. As a result there is spatial autocorrelation in the model. One method to test for the presence of spatial autocorrelation is Morans I (Moran, 1948).

$$I = \frac{n}{S_0} \frac{\sum_i \sum_j w_{i,j}(Y_i - \bar{Y})(Y_j - \bar{Y})}{\sum_i (Y_i - \bar{Y})^2}$$

Where n is the number of spatial units i and j, Y is the dependent variable and $\bar{Y}$ the population mean of Y, $w_{ij}$ is the spatial weights matrix and $S_0$ is the aggregate of all spatial weights. With the global Morans I we test the null hypothesis that the dependent variable is randomly distributed across space. After running a Morans I test in R, we get a p-value such that we reject the null hypothesis that our data is not spatially clustered.

One initial method of testing for spatial dependence is to use a spatial lag Spatial Autoregressive (SAR) or spatial error SAR. A spatial lag uses a weighted average of neigbouring values where neighbours are defined using a spatial weights matrix. Neighbours can be defined using contiguity (whether they share a border) or by distance or k nearest neighbours. Weights are row standardised so that overemphasis is not placed on areas with many neighbours.

$$y = \rho W_y + X\beta + \varepsilon$$

Where w is the spatial weights matrix. While the spatial error model is:

$$y = X\beta + u$$

$$u = \lambda W u + \varepsilon$$

(Bivand et al., 2008)

The main difference between a spatial lag and spatial error model is in the autocorrelation that exists. If the spatial autocorrelation is between the residuals, a spatial error model may be used, if spatial autocorrelation exists in the dependent variable we use a spatial lag model, that is the Y in one location i is being influenced by the Y in another location j. We choose the model with the highest log likelihood and $r^2$ value which is in our case is the spatial error model.

Given the results of the Geary's C test, there is spatial autocorrelation in our data, which will bias our results. A correction must be applied which controls for the spatial dependence in our model. One method is to run a spatial error model. We use the same dependent and independent variables as the OLS except we now include a spatial weights matrix. The spatial weights matrix is calculated using queen contiguity with row standardised weights (that is the weights in each row of the matrix sum to one, this helps in not overemphasise areas, which have many neighbours). Figure 3, examines the AIC from all of the error and lag models, the spatial error is the best model as it has the lowest AIC value. A spatial error model is performed using k-nn.

**Figure 3: AIC of OLS, spatial lag, spatial error and Queen lag models**

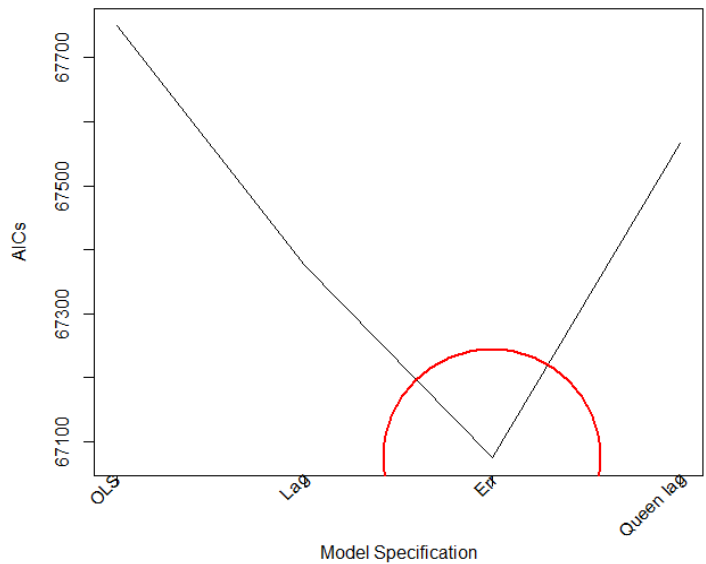

Table 5, shows the results of the spatial error model. Comparing the spatial error model to the OLS (table 3), we see a change in the magnitude of the coefficients. The overall fit of the model represented by the R2 has improved. In addition our residuals were not independent as there was spatial dependence in the data. After running the spatial error model. The coefficient sign on the variables are to be expected. Working in commerce, having internet and high education all having a positive impact on MGHDI.

**Table 5: Spatial error model results**

```
Spatial Error Model - K-nn (n=23)
=============================================================
                                   Dependent variable:
                          ----------------------------------
                          Median Household Gross Disposable Income
-------------------------------------------------------------
Age 65+                              -530.684***
                                        22.864

Born outside Ireland                 -465.625***
                                        15.569

Single Person                        -410.676***
                                        29.081

Lone Parent                          -149.039***
                                        13.321

Unemployed                           -483.272***
                                        44.586

Commerce workers                      282.154***
                                        20.818

Internet (Broadband)                  194.175***
                                        17.674

Low Education                         -67.870**
                                        25.004

High Education                        581.559***
                                        17.606

Constant                            37740.470***
-------------------------------------------------------------
Observations                           3,409
Log Likelihood                       -33,936.75
Lambda: -0.0018631, LR test value: 0.49511, p-value: 0.48166
AIC                                   67,897
=============================================================
Note:                         *p<0.1; **p<0.05; ***p<0.01
```

Comparing the result of the spatial error model to the OLS, we can see that there was no change in the direction of the coefficients, only in the magnitude. Lambda, an indicator of spatial autocorrelation is highly significant. The sign on the various independent variable are expected. We also examine the difference in residuals by subtracting the residuals from the OLS model from those of the spatial error model. The results show that there is greater unexplained heterogeneity in the spatial lag model with the exception of a few areas (North-west and South East). Once we have specified our model, we firstly examine the spatial distribution of income. Local Morans I is utilised to highlight hotspots and coldspots, areas with high levels of spatial autocorrelation. Finally, the results of the GWR model are presented.

*Spatial Regression - GWR*

In order to explore the spatial heterogeneity GWR is utilized. Comber et al., (2020) is a go to resource to help in deciding whether a GWR model is appropriate. The choice of model is determined by the bandwidth. The relationship between variables maybe considered local or global. If the bandwidth is ~80% of the maximum distance, the relationship can be considered global and GWR may not be appropriate.

Given the existence of spatial dependence and after examining the fixed bandwidth results between variables it seems appropriate to utilize GWR. The first step involves the creation of the spatial weights matrix. This determines number of observations included in each local model. A weights matrix is created using either contiguity (areas which border/touch each other), K-nearest neighbours (k-nn) (selecting the n nearest neighbours) or using a fixed distance band (a 5km buffer zone around the area of interest). Given the variation in the size of EDs, a fixed bandwidth is not appropriate. A fixed bandwidth would include more observations when examining urban areas compared to rural areas. Given the nature of EDs which vary from small city centre areas to large rural areas, K-nn is an adaptive bandwidth method. The same number of EDs will be under the kernel each time, where EDs are more closely packed together the radius will be lower than where the EDs are large (as in rural areas). This is deemed the most appropriate method as it can cope with both urban and rural areas. The optimal number of k-nn is chosen by minimising the AIC value after multiple iterations. A variable bandwidth is used to account for these differences. The variable bandwidth method uses k nearest neighbors to specify the ideal bandwidth by minimizing the AIC value.

$$w_{ij} = 1 \ \forall i,j$$

In a global OLS regression, the weighting is given by:

$$w_{ij} = exp\left[-\frac{1}{2}(d_{ij}/b)^2\right]$$

Where b is the chosen bandwidth, and d is the distance between point i and j. The weighting corresponding to where i and j are the same point will be unity, 1. As the distance between i and j increases, the weighting used decreases according to a Gaussian curve (Fotheringham et al., 2002). The bandwidth determines the distance at which $w_{ij} = 0$.

Given the heterogeneity between regions, difficulties in fitting a linear function to a highly non-linear function emerge, GWR attempts to overcome this problem by providing a nonparametric estimate of β for each location j (Brunsdon et al., 1998) so that we get the equation:

$$y_i = \sum_j X_{ij}\beta_j(p_i) + \varepsilon_i \quad \text{(Brunsdon et al., 1998)}$$

Where $p_i$ is the location of i. Geographically Weighted Regression (GWR) (Brunsdon et al., 1998) can be used to model spatial heterogeneity in relationships by explicitly considering the spatial structure of the study area. Separate estimations of the parameters are made for each data location. This method moves away from a whole map approach which examines averages across space, towards a measuring of the spatial variation in the relationships (Brunsdon et al., 1998).

$$y_i = \beta_0(u_i, v_i) + \sum_k \beta_k(u_i, v_i) x_{ik} + \varepsilon_i$$

Where $(u_i, v_i)$ are the coordinates of the ith point in space.

Despite multicollinearity not being present in the global model as evidenced by the variables having a low VIF score, despite this variables may have a high VIF score at the local level. We follow the methodology outlined in Gollini et al., (2013) in how to deal with local multicollinearity. We start by performing the BKW method (Belsley et al., 1980) to test for the presence of multicollinearity. BKW method can identify near dependencies in data. A condition index score of 68.7 is an indication that multicollinearity is a problem with our model given the current specification and variable choices. Using the local condition number we can identify areas where local multicollinearity is high and decide how to remedy the problem. We run the basic GWR model (Lu et al., 2019) using an adaptive bandwidth which uses k-nearest neighbours (k-nn). Given the variability in the area of EDs, the advantage of k-nn is that it ensures we have a consistent number of observations in our local model compared to if we used a fixed bandwidth (fixed distance). The optimal number of neighbours is selected by minimising the coefficient variation (CV). Using the variable bandwidth, the GWR basic is performed and the local condition number is calculated for each ED. From figure 4 we can see that the local condition numbers were higher before the variable internet was excluded. After removing internet local multicollinearity remains an outside of cities in the commuter belt.

**Figure 4: Spatial distribution of local condition numbers**

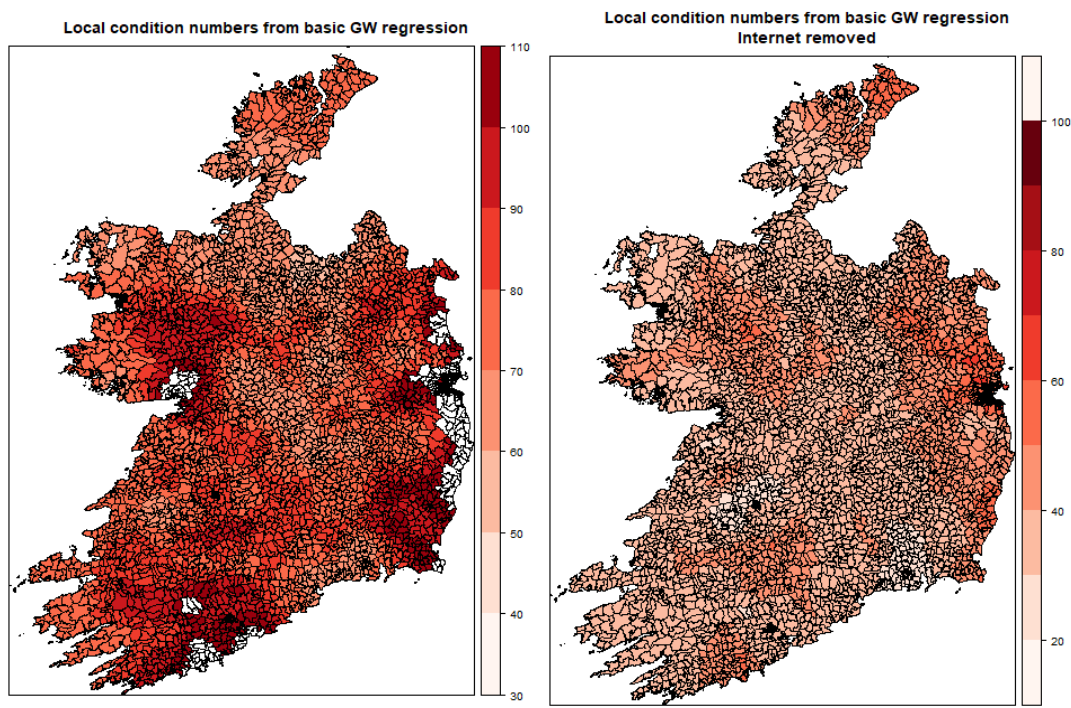

Given the high local condition number in some EDs, local multicollinearity is an issue. One remedy for the issue to drop variables from the analysis. After dropping the share of households with internet, the local condition numbers are recalculated for the model. As we can see from figure 5, the local condition numbers remain high. There are a number of remedies for this issue, continue dropping variables until the condition numbers are in an acceptable range or use Geographically Weighted Ridge Regression (Wheeler, 2007) applying a local compensation for areas where condition numbers are high and local multicollinearity is an issue. First using the CV score, we calculate the range of local condition numbers for a range of models using different variable selections. We can observe the CV locally and try and minimise or ensure no ED has a score above 30 in the local model. All models will have multicollinearity present however we can control for this as best we can (Gollini et al., 2013). Following the procedure outlined in (Gollini et al., 2013) we calculate the range of local condition values for a number of model specifications. From this we can decide which model is best of accounting for local multicollinearity. A number of options of variable, we may decide to only drop three to four variables in which case local multicollinearity will still be an issue in some areas. To account for this geographically weighted ridge regression can be used where a local ridge is added in the event the local condition value exceeds a user specified threshold, 30 for example. The second option is to use the model including only single households and high education variables in which case the local condition values are below the threshold of 30 and a local ridge is not required.

**Figure 5: Local condition values by model specification**

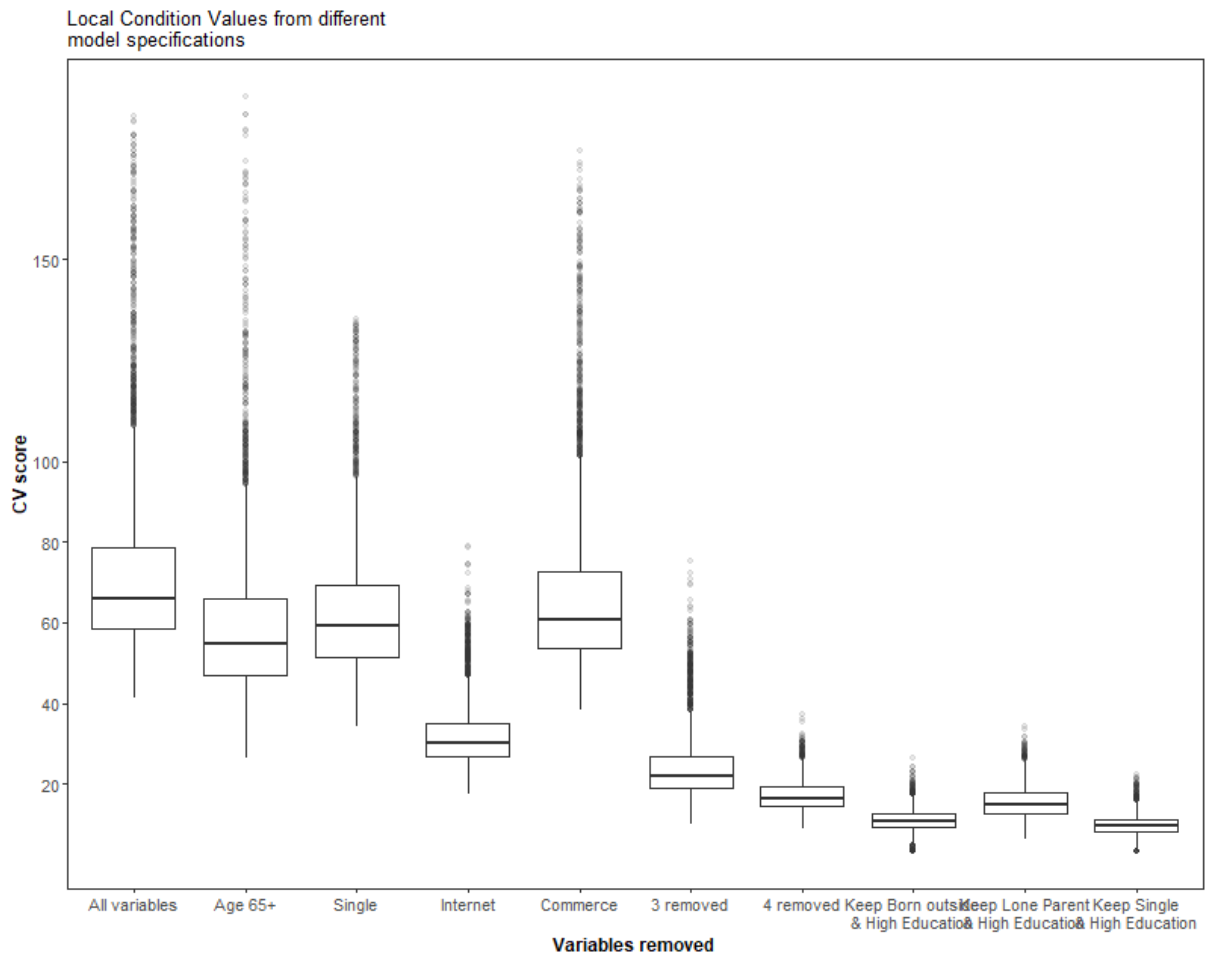

Unlike previous GWR models considered which use the same fixed or variable bandwidth for all variables, a multi-scale GWR model can specify different bandwidths for each variable in the model (Li and Fotheringham, 2020; Lu et al., 2017, 2018; Yang, 2014). Table 6 shows the optimal bandwidth for each variable. We interpret the bandwidths in relation to the maximum bandwidth, that is the largest distance between two EDs (Comber et al., 2020). The bandwidths for age 65+, born outside Ireland and commerce can be considered global as they tend towards the maximum bandwidth distance. The remaining variables can be considered local as the bandwidths are small, relative to the maximum bandwidth. Although the optimal bandwidth for the internet variable is local, we discovered from that local multi-collinearity becomes a problem when we include it in our model. Also by including age 65+, born outside Ireland and commerce we do not learn anything about the spatial heterogeneity of income given the global bandwidth.

**Table 6: Variable optimal bandwidth**

| Variable | MGHDI | age 65+ | born outside Ireland | single person | lone parent | commerce | internet | high education |
|---|---|---|---|---|---|---|---|---|
| Bandwidth | 24,703 | 461,307 | 461,227 | 124,321 | 50,747 | 461,276 | 89,554 | 89,345 |
| Share of max bw | 0.05 | 1.00 | 1.00 | 0.27 | 0.11 | 1.00 | 0.19 | 0.19 |
| Max bw | 461,356 (meters) | | | | | | | |

The GWR-LCR approach which uses a local ridge adjustment as opposed to GWRR where a ridge adjustment is applied to all areas not just those where local multicollinearity is an issue. We generate three GWR models; Model 1) using a reduced number of variables MGHDI, single person, high education with the basic GWR with an adaptive bandwidth Model 2) only removing the internet variable and including all MGHDI, single person, high education, age 65+, born outside Ireland, lone parent, commerce and no education, using a LCR-GWR with an adaptive bandwidth and Model 3) using MS-GWR using a reduced number of variables MGHDI, single person, lone parent and high education with variable specific bandwidths. The Akaike information criterion (AIC) or corrected AIC (Akaike, 1973) of all models are consulted (figure 6). We also include the AIC value of the OLS and Spatial Error models previously discussed. AIC is considered a measure of the information distance between the model distribution and the 'true' distribution (Fotheringham et al., 2002). Model 1 gives an AICc of 67706.09, Model 2 an AICc of 67916.4 and Model 3 an AICc of 67639.68. We proceed with the model with the lower AIC value Model 3 which uses MS-GWR and proceed.

**Figure 6: AICc value of model specifications**

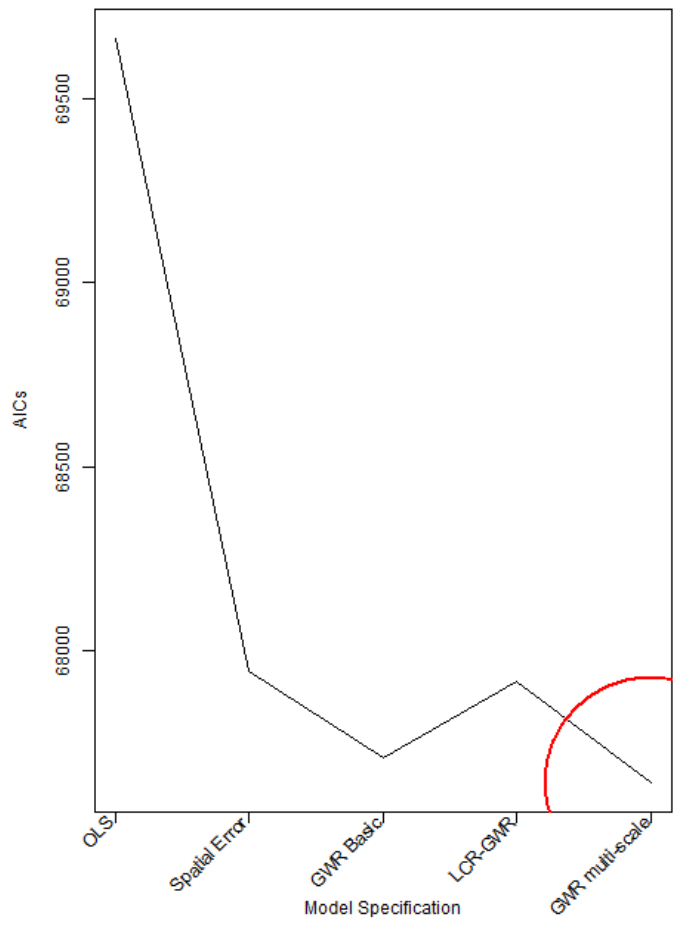

# Results

*Results I: Spatial Distributions*

The first set of results examine the spatial distribution of income and also examine reliance on social welfare and proportion of medical cards to examine any differences between the different measures. Spatial analysis is also carried out on MGHDI and Anselin local Moran's I is created to identify statistically significant pockets of high/low income.

**Figure 7: Median gross household disposable income (MGHDI) by electoral division (ED) using jenks breaks**

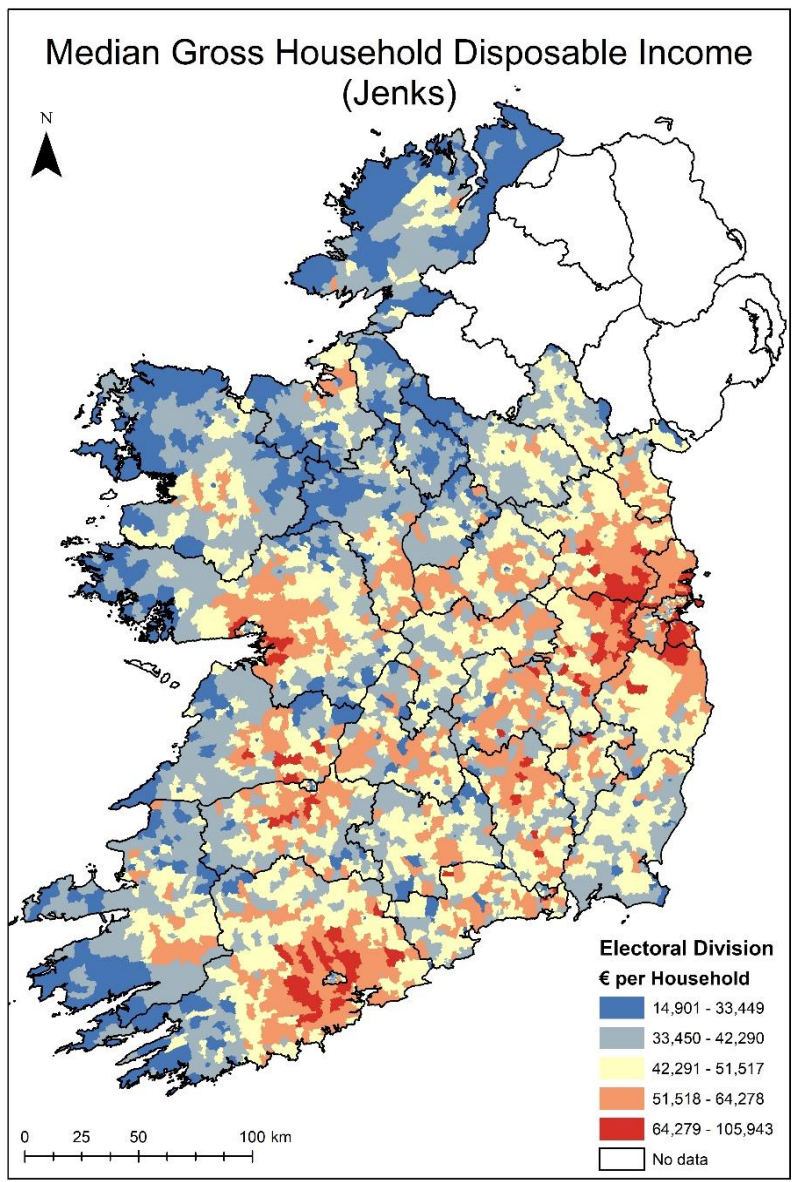

From figure 7 we can see the highest levels of MGHDI are located around the two major cities of Cork and Dublin. Higher incomes in cities is not surprising, and the explanations go back to the work of Marshall (Marshall, 1890) who discussed concentration of industry and the benefits they receive from knowledge spillovers and hence greater productivity. Dense labour markets in cities attracts both firms and workers. can also have highly educated workforce and rich human capital which benefits growth (Romer, 1990). This urban wage premium may exist due to knowledge accumulation by the labour market (Glaeser and Mare, 2001). There is also the idea there is a wage premium to living in cities that have a high amenity value and quality of life (Glaeser et al., 2001). Within a city, the areas with the high levels of amenity value are likely to attract the highest earners and hence the urban rents here will be high (Brueckner et al., 1999).

**Figure 8: Percentage of households with social welfare as main source of income**

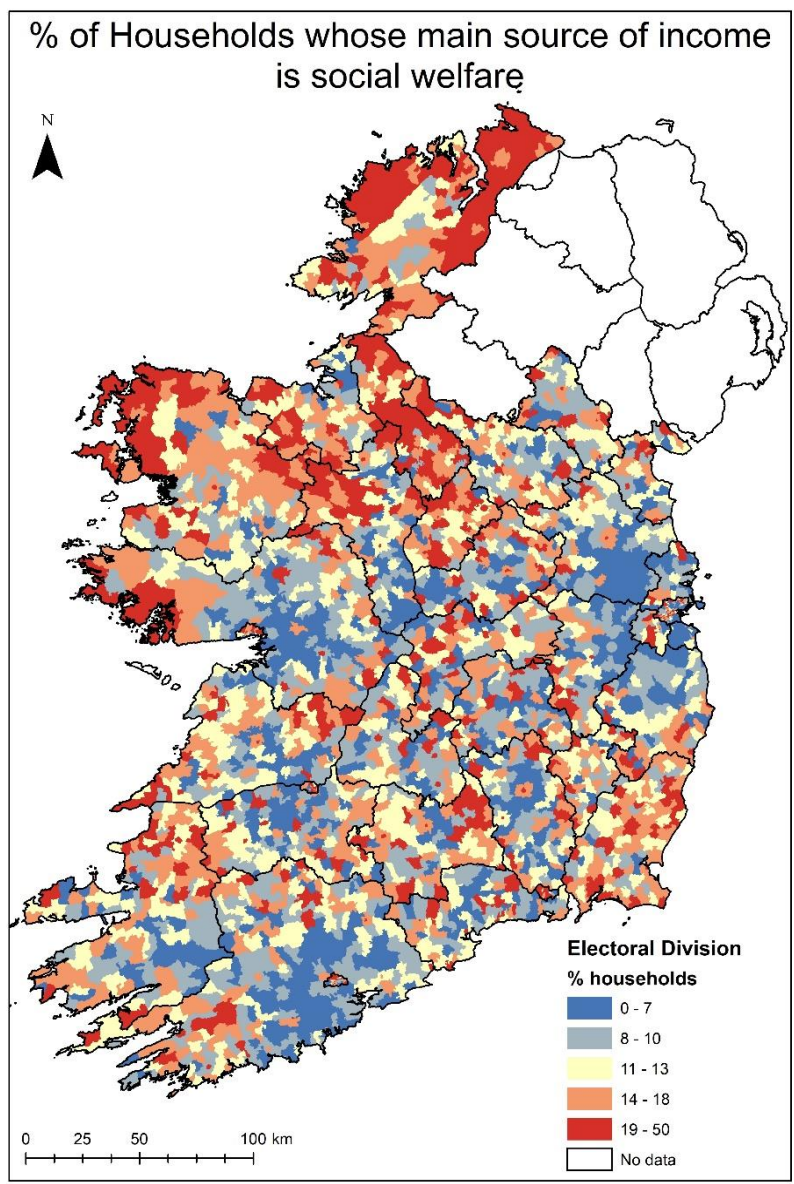

In figure 8, we examining the distribution of social welfare payments, rural households appear to have higher levels of reliance on social welfare transfers. In some areas, the percentage of households whose main source of income is social welfare is as high as 50% in some areas. Referring to the distribution of household incomes, there is obviously a link between the two maps. Those areas with lower levels of income have a higher level of households who are dependent on social welfare payments.

**Figure 9: Percentage of households with a medical card**

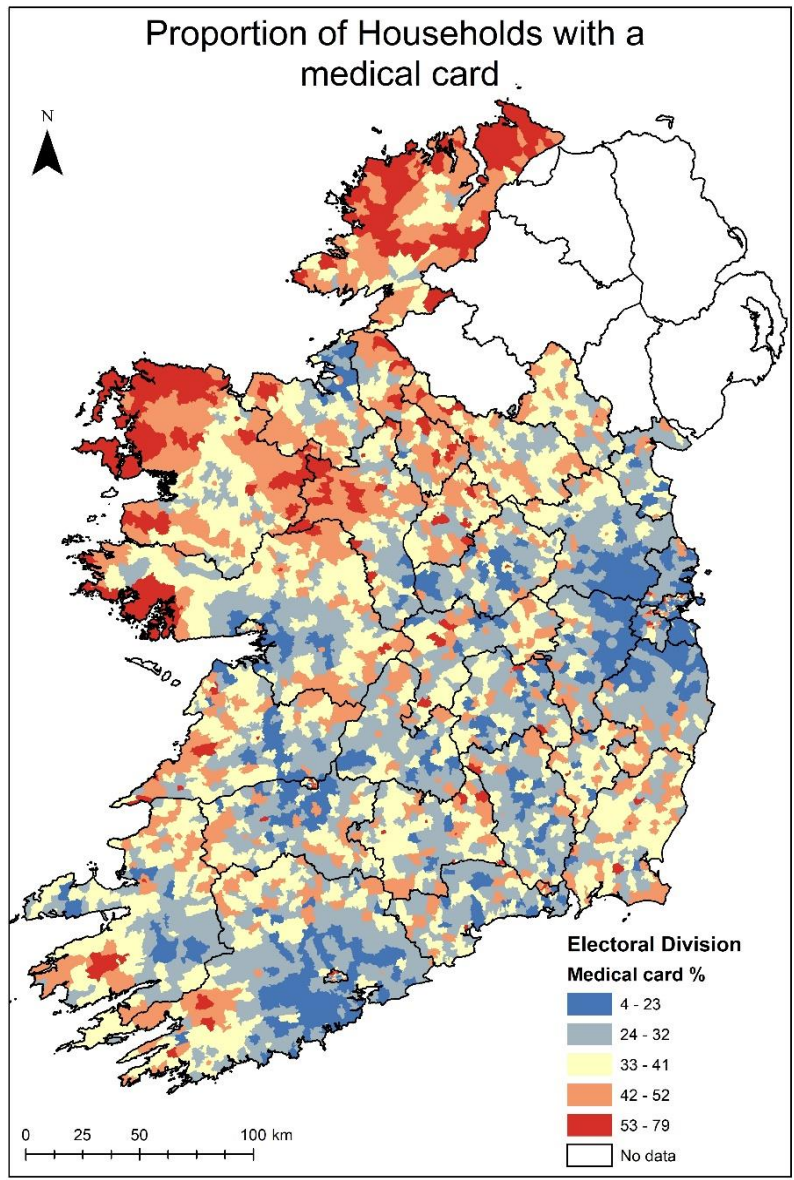

Figure 9 shows proportion of households with a medical card. Since the medical card is means tested, only those households below the threshold will qualify. The medical card is also available to old age pensioner, which may be an explanation behind the high level of households in the north-west. These areas also have a high level of old age dependency.

**Figure 10: Anselin Local Moran's I of median gross household disposable income**

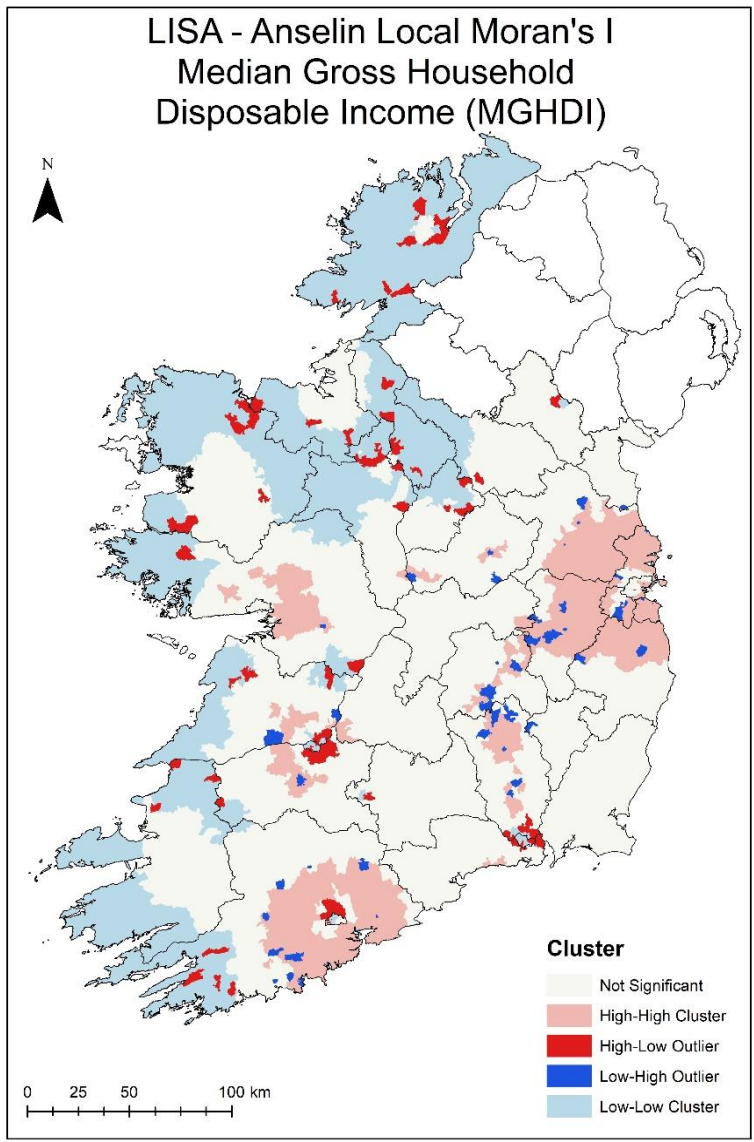

Using local Moran's I we can identify clusters of high/low income. There is clearly and urban/rural divide with higher income areas located in the city commuter belts. A centre/suburb pattern also emerges. The high areas are mostly around the commuter belts of Cork and Dublin and into periphery areas. There are smaller areas of high income around Galway, Limerick and Kilkenny. Low levels of income are located in the south-west, north-west and border areas. The high-low and low-high are useful clusters to identify pockets of deprivation or influence in an otherwise rich/poor region. Local Moran's I identifies areas that are statistically different, It is a data driven method of identifying patterns without relying upon human interpretation which can be influenced by bias.

*Results II: GWR results – Drivers of income*

GWR enables us to explore the difference in the relationship between the dependent and independent variables across space. GWR will present us with a coefficient for each variable allowing us to show how the magnitude in the independent changes depending upon the area. For example, in some areas high education will have a greater effect on MGHDI compared to others.

**Table 7: MS-GWR Results**

```
***********************************************************************
            GWR with Parameter-Specific Distance Metrics
***********************************************************************

*********************Model calibration information*********************
 Kernel function: Bisquare
 Fixed bandwidths for each coefficient:
            (Intercept)      Single    Lone P. High Educ.
 Bandwidth       17780        90568      36133     101899

*****************Summary of GWR coefficient estimates:*****************
              Min.    1st Qu.   Median    3rd Qu.     Max.
 Intercept  21474.33 41220.93 46601.10 49550.03 64718.394
 Single      -944.33  -799.05  -705.97  -594.61  -357.027
 Lone P.     -759.92  -477.57  -379.62  -314.99    98.586
 High Educ.   290.78   496.34   598.63   665.71   673.195
**************************Diagnostic information***********************
 Residual sum of squares:   71652347924
 R-square value:            0.8372268
 Adjusted R-square value:   0.8150975
 AICc value:                67639.68
```

*The* GWR results in table 7 summarise the variation in the coefficients by area. The table highlights the heterogeneity. For some areas the independent variables will have a higher/lower positive/negative influence on income. The following maps show how the parameter values vary across space from the MS-GWR model. The magnitude of the relationship between the dependent and independent variable varies across space. From the results we can see how the drivers of income in Ireland are not uniform across space. Increases in educational attainment have a stronger positive influence on changes in income compared with the being a lone parent or single person household. Given the higher R2, the MS-GWR explains more variation compared to a global OLS model. The signs of the parameters are expected. Single person household is mostly negative and high education is positive. The advantage of using GWR is we can examine how these parameter values vary across space. Given the multi-scale bandwidths used, (24km for the intercept, 124km for single person, 36km for lone parent and 89km for education), the value range of the parameters may be similar by region.

**Figure 11: GWR parameter values - Intercept**

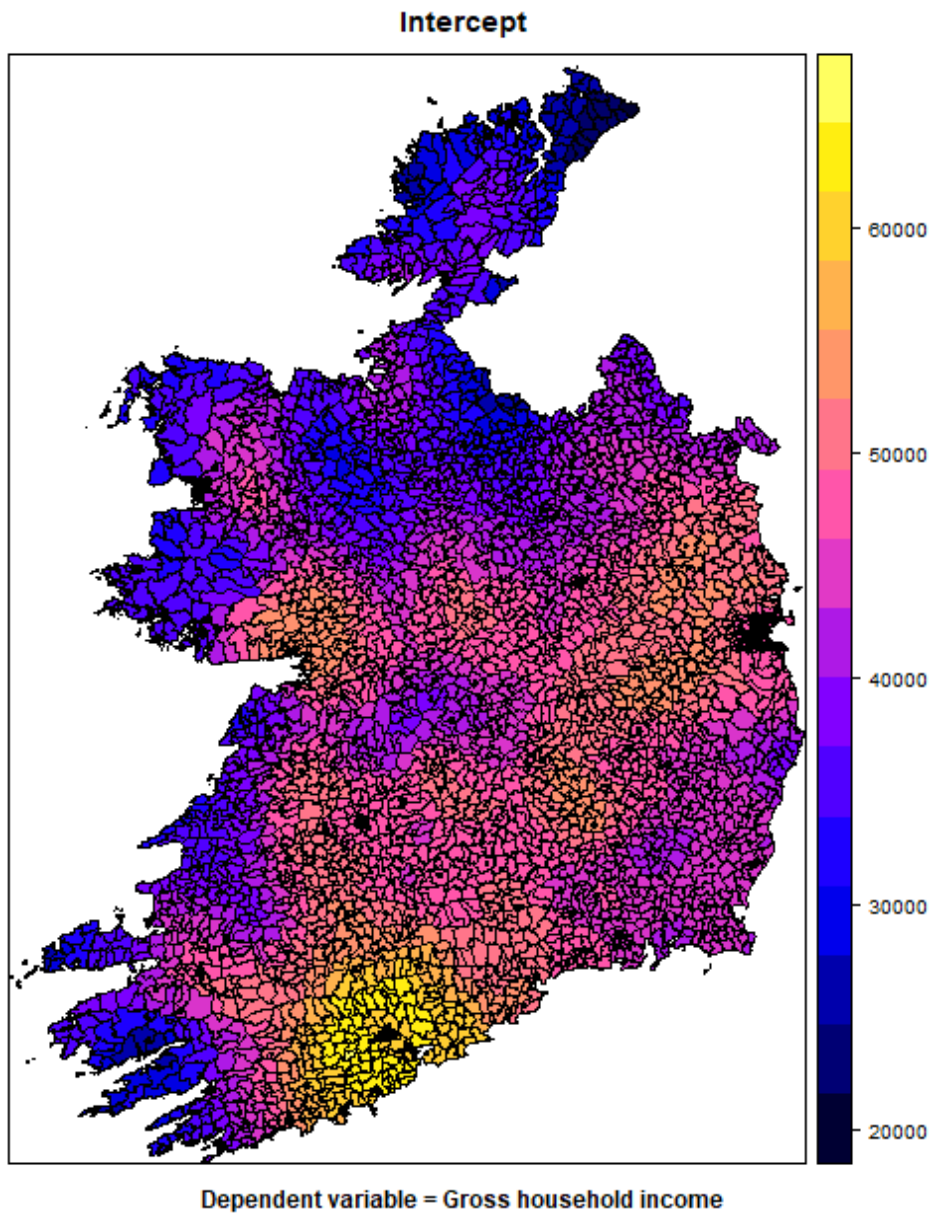

Figure 11 shows the intercept and highlights which areas our model predicts would have the highest levels of income if our independent variables are zero. The highest values are located around Cork city and not Dublin, which suggests our explanatory variables have more of a positive or less of a negative impact on Dublin compared to Cork. Like in figure10, the lowest incomes are located in the north-west.

**Figure 12: GWR parameter values - Single person household**

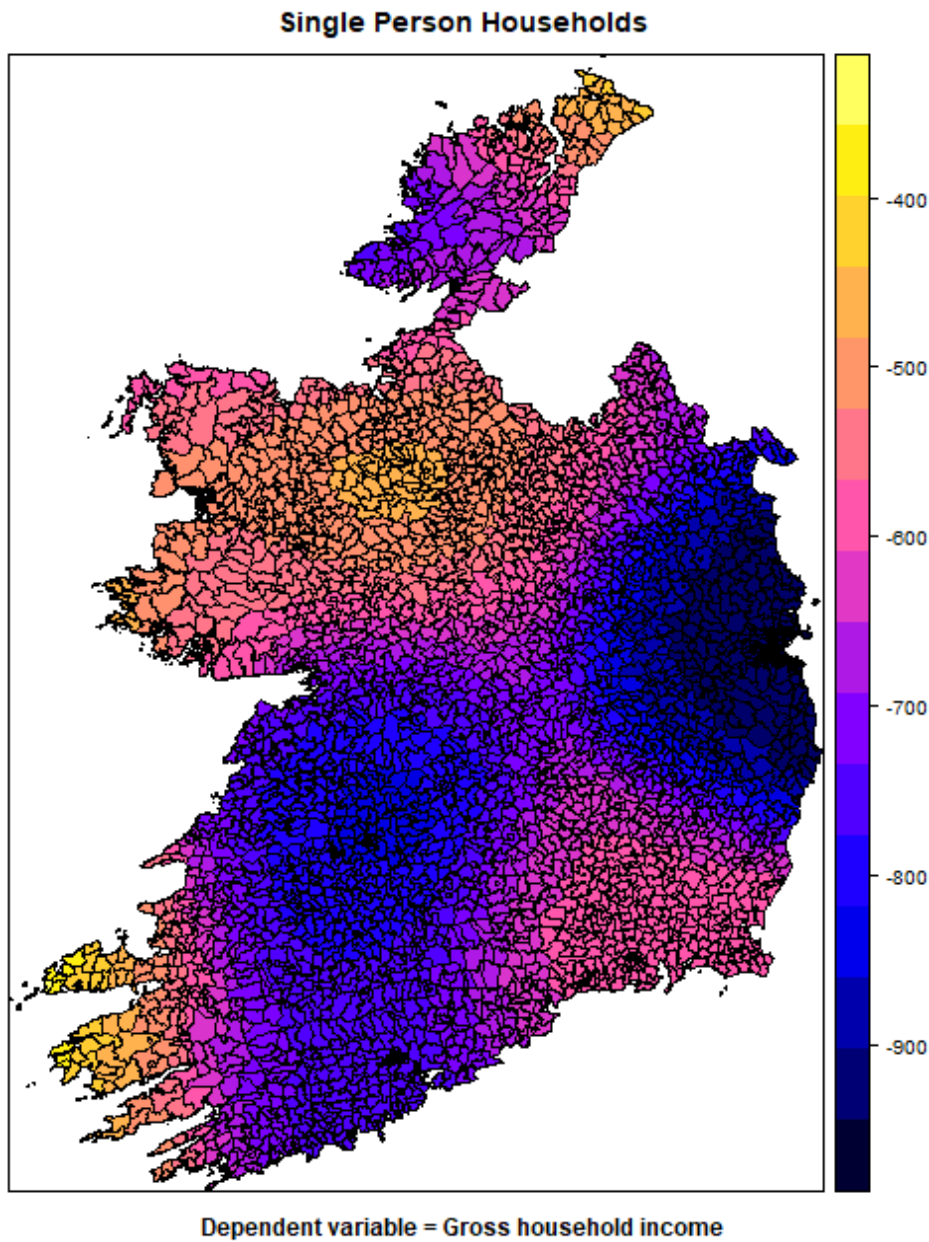

Figure 12 shows that Single person households are associated with lower household income. This is not surprising as these are also single earner households. The lowest parameter values for single person households are found in the east around Dublin and around the cities of Limerick and Cork. In the commuter regions of Limerick, Cork, Dublin and to a lesser degree Galway, increases in the share of single person households have stronger negative changes on income. The stronger negative change is observed in the Dublin commuter belt. The pattern suggests that single earner households in Dublin earn on average less income, compared to single person households in Galway, Cork and Limerick. Given the data is at the ED level, we are unable to examine what the characteristics of a 'typical' single person household is in Dublin.

**Figure 13: GWR parameter values - High Education**

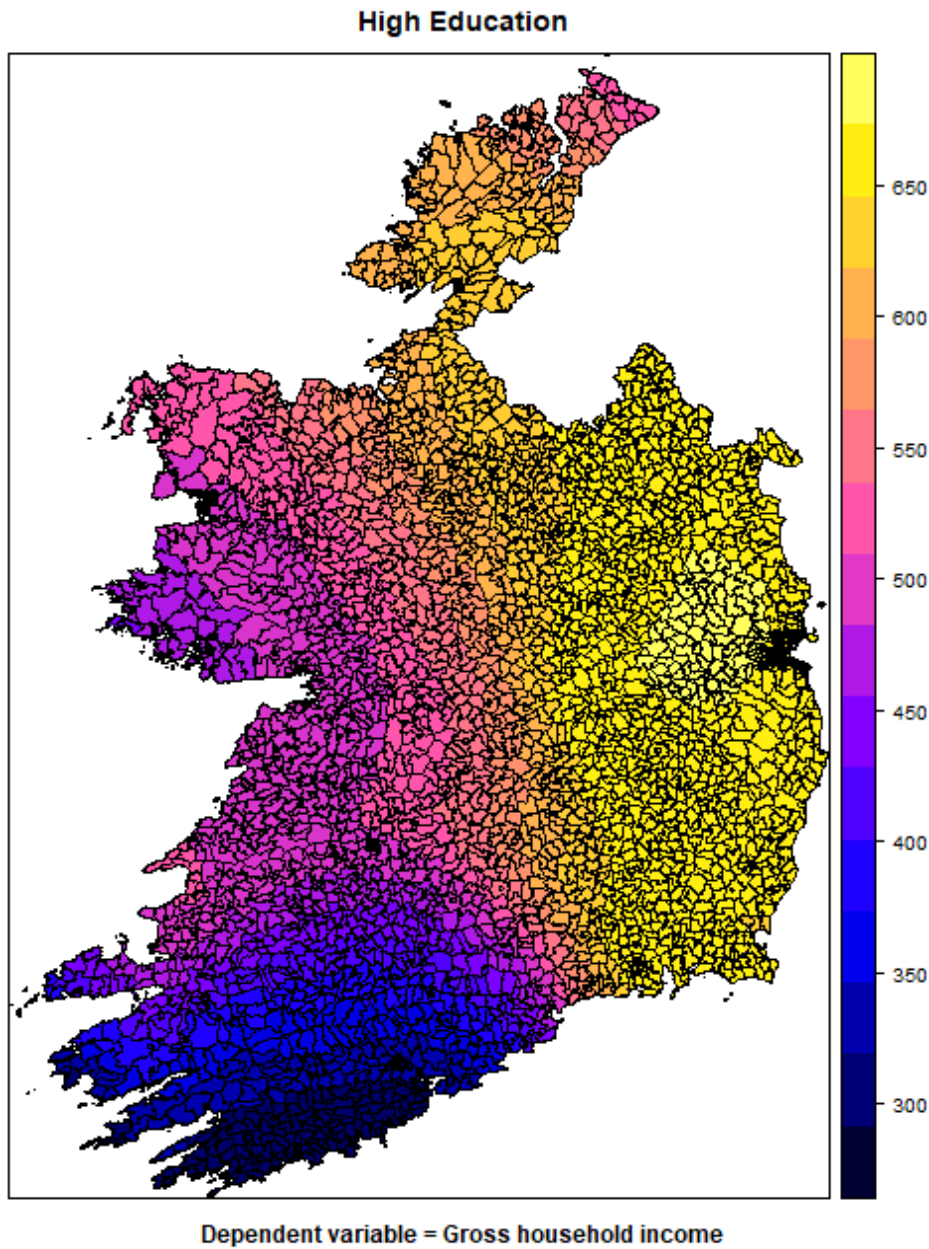

Figure 13, shows the influence of high education on income. The parameters for high education show an east-west difference with the highest parameter values in the east around Dublin and the lowest in the south east. Increases in educational attainment around Dublin have a stronger positive influence on changes in income compared to areas in Galway, Cork and Limerick. As you move outside of the Dublin catchment area, high education has less of a positive influence on income. This is to be expected from theory, where cities benefit from pooled labour markets and are attractive to high education individuals. The low influence of education around Cork city is surprising. One explanation is that areas around Cork city have more of a mix of households compared to Dublin. High earning households may earn their income in industries where high educational attainment is not an important requirement.

**Figure 14: GWR parameter values - Lone Parent**

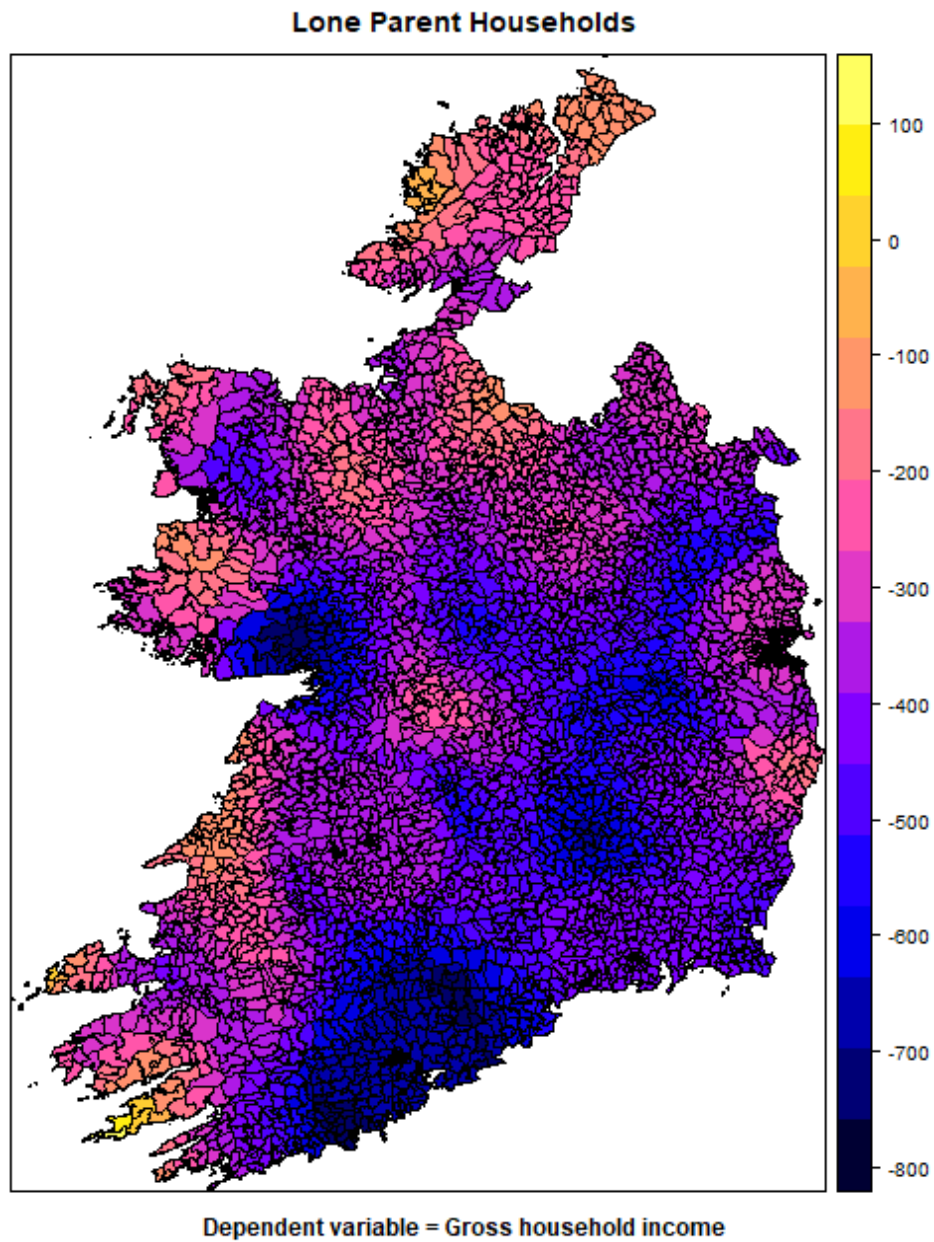

In figure 14 the low bandwidth for lone parent becomes apparent. The bandwidth for lone parent is the lowest of the four independent variables at 36km. The parameter values will therefore will more localised. Overall the influence of lone parent households on income is largely negative, except in a few remote areas in Kerry and Donegal. The influence on income is highly negative in areas around Cork and Galway. Share of lone parent households is also negative around Dublin and Limerick but to a lesser extent. The differences between the four main cities is interesting and highlights the local nature of income. The high negative influence of lone parent households around Cork and Galway warrants further investigation.

**Conclusion**

Examining the CSO electoral division income data using a spatial analysis approach, one can identify on a national level the drivers of income and at a local level the differential impact of these drivers. Adopting a spatial approach, gives us a greater understanding of the current economic situation in Ireland at a local level. The sophisticated methodologies employed give a greater understanding of the spatial relationships that exist. The relationships between income and socio-demographic characteristics are complex. This paper has contribute to this research by presenting a complete picture of the Irish spatial distribution of income at an ED level.

An examining of the spatial distribution of median disposable income highlighted the disparities, firstly between urban and rural areas and secondly between the cities of Dublin and Cork and other cities. This pattern was further highlighted using Anselin local Moran's I with hotspots of high-high areas around the main urban areas but interestingly not within. Commuter areas have on average higher incomes compared to core city centre locations. Further research should examine whether Irish city centre locations are more in line with Detroit or Paris and attempt to explain these differences paying particularly attention to amenities.

The social welfare and medical card data highlights particularly low levels of income in the north west of the country. Given urban areas have higher levels of income it is not surprising to find the lowest levels of reliance on social welfare there. There are however pockets of high reliance even within cities which can be utilised to identify areas of deprivation.

Using the MS-GWR enabled us to overcome issues of local multicollinearity and to explore the spatial heterogeneity of median gross household disposable income. The results from GWR show found that the share of lone parent households has less of an influence on income in Dublin compared to other variables such as high education attainment and share of single person households. High education attainment has a higher positive influence on income around Dublin that it has in other parts of the country, this could be due to the higher share of professional and technical jobs in Dublin compared to other areas. The influence of independent variables also varied by city, highlighting the heterogeneity between Irish cities.

The results highlight the differences at a regional level in income and how its influences vary across space. If a Global OLS model is used, we incorrectly make the assumption, the influence of high education on income is homogenous across space. This is clearly not the case as high education and income exhibit spatial dependence as shown by the Geary's C measure.

Future work should expand upon this analysis by including additional drivers related to housing and commuting. This could help to inform policy around how household income, housing and commuting are all interacting. Given the hedonic nature of housing, prices vary across space. Lower house prices in periphery areas makes larger houses more attainable but at the same time can increase both journey time and cost.

**Abbreviations**

| | |
|---|---|
| COPA | census of population analysis dataset |
| CSO | central statistics office |
| ED | electoral division |
| GDA | greater Dublin area |
| GWR | geographically weighted regression |
| GWRR | geographically weighted ridge regression |
| k-nn | k nearest neighbours |
| MAUP | modifiable areal unit problem |
| MGHDI | median gross household disposable income |
| MSE | mean square error |
| MS-GWR | multi-scale geographically weighted regression |
| PDI | pobal deprivation index |
| PIR | personal income register |
| SAPS | small area population statistics |
| SAR | spatial autoregressive |
| VIF | variance inflation factor |